\documentclass{osa-article}

\journal{oe}
\graphicspath{{Images/}}

\newcommand{\quarterrange}{90$\pm$6}
\newcommand{\fullrange}{120$\pm$8}
\newcommand{\staticrange}{200}

\usepackage{mathrsfs}
\date{\today}

\articletype{Research Article}
\usepackage{lineno}
\begin{document}
\title{Piezo-deformable Mirrors for Active Mode Matching in Advanced LIGO}

\author{Varun Srivastava, \authormark{1,*} 
Georgia Mansell, \authormark{2,3}
Camille Makarem, \authormark{4}
Minkyun Noh, \authormark{5}
Richard Abbott, \authormark{4}
Stefan Ballmer, \authormark{1}
GariLynn Billingsley, \authormark{4}
Aidan Brooks, \authormark{4}
Huy Tuong Cao, \authormark{6}
Peter Fritschel, \authormark{2}
Don Griffith, \authormark{4}
Wenxuan Jia, \authormark{2}
Marie Kasprzack, \authormark{2}
Myron MacInnis, \authormark{2}
Sebastian Ng, \authormark{6}
Luis Sanchez, \authormark{4}
Calum Torrie, \authormark{4}
Peter Veitch, \authormark{6}
Fabrice Matichard, \authormark{2, +}}

\address{\authormark{1}Department of Physics, Syracuse University, Syracuse, New York 13244, USA\\
\authormark{2}LIGO Laboratory, Massachusetts Institute of Technology, Cambridge, Massachusetts 02139, USA\\
\authormark{3}LIGO Hanford Observatory, Richland, Washington 99352, USA\\
\authormark{4}LIGO Laboratory, California Institute of Technology, Pasadena, CA 91125, USA\\
\authormark{5}Department of Mechanical Engineering, The University of British Columbia, Vancouver, BC V6T 1Z4, Canada\\
\authormark{6}OzGrav, University of Adelaide, Adelaide, South Australia 5005, Australia}
\email{\authormark{*}vasrivas@syr.edu \authormark{+}fmatichard@lbl.gov} %% email address is required


\begin{abstract}
The detectors of the laser interferometer gravitational-wave observatory (LIGO) are broadly limited by the quantum noise and rely on the injection of squeezed states of light to achieve their full sensitivity.
Squeezing improvement is limited by mode mismatch between the elements of the squeezer and the interferometer.
In the current LIGO detectors, there is no way to actively mitigate this mode mismatch.
This paper presents a new deformable mirror for wavefront control that meets the active mode matching requirements of advanced LIGO.
The active element is a piezo-electric transducer, which actuates on the radius of curvature of a 5~mm thick mirror via an axisymmetric flexure.
The operating range of the deformable mirror is \fullrange~mD in vacuum and an additional \staticrange~mD adjustment range accessible out of vacuum.
Combining the operating range and the adjustable static offset, it is possible to deform a flat mirror from -65~mD to -385~mD.
The measured bandwidth of the actuator and driver electronics is 6.8 Hz.
The scattering into higher-order modes is measured to be $<$0.2$\%$ over the nominal beam radius.
These piezo-deformable mirrors meet the stringent noise and vacuum requirements of advanced LIGO and will be used for the next observing run (O4) to control the mode-matching between the squeezer and the interferometer.
\end{abstract}

%%%%%%%%%%%%%%%%%%%%%%%%%%%%%%%%%%%%%%%%%%%%%%%%%%%%%%%%%%%%%%%%%%%%%%%%%%%%%%%%%%%%%%%%%%%%%%%%%%%%%%
%%%% Section I
%%%%%%%%%%%%%%%%%%%%%%%%%%%%%%%%%%%%%%%%%%%%%%%%%%%%%%%%%%%%%%%%%%%%%%%%%%%%%%%%%%%%%%%%%%%%%%%%%%%%%%
\section{Introduction}\label{sec:intro}
In September 2015 the Advanced LIGO (aLIGO) detectors \cite{aLIGO2015} made the first direct observation of gravitational waves from a binary black hole merger \cite{GW150914}.
This first detection kicked off the exciting new field of gravitational-wave astronomy.
Since the first detection, aLIGO, together with the Advanced Virgo observatory \cite{Acernese_2014}, has undergone incremental upgrades to improve the sensitivity to gravitational waves, and subsequently observed 90 gravitational-wave events \cite{GWTC1,GWTC2,GWTC3} from binary black holes~\cite{GW150914}, binary neutron stars~\cite{GW170817, GW190425}, and neutron-star black-hole binaries~\cite{GWNSBH}. 

The aLIGO detectors are two dual-recycled Fabry-P\'{e}rot Michelson interferometers located in Hanford, Washington and Livingston, Louisiana.
Incoming gravitational waves cause a minuscule displacement of the test masses - 40~kg mirrors which make up the 4km long arms of the Michelson interferometer.
The gravitational-wave readout is a measure of the differential arm length of the interferometer.
The sensitivity of the current gravitational-wave detectors is broadly limited by quantum noise~\cite{O3comm}.
Below 50~Hz, the quantum noise manifests itself as radiation pressure noise, as photons circulating in the arm Fabry-P\'{e}rot cavities impart momentum to the test masses. 
Above 100~Hz, the sensitivity of the detector is limited by quantum shot noise.
The sensitivity of the aLIGO detectors is improved by the injection of squeezed states of light.
In the most recent observing run (O3) frequency-independent squeezing was injected into the aLIGO detectors, reducing quantum shot noise by roughly $3~\mathrm{dB}$ compared to when no squeezed light is injected \cite{Tse_2019}.

The aLIGO detectors are currently being upgraded. 
One of the major upgrades for the next observing run (O4) is the implementation frequency-dependent squeezing.
Frequency-dependent squeezing is achieved by reflecting squeezed light off a long-baseline filter cavity, with the filter cavity pole at the desired rotation frequency.
For O4, a 297~m filter cavity will be installed at each of the LIGO sites.
To maximize the squeezing improvement to detector sensitivity, the mode-matching losses between the various optical cavities need to be minimized.
The negative effect of mode mismatch on squeezed photons is twofold: mode mismatch causes optical loss and adds phase noise to the squeezed beam~\cite{McCuller2021}.
For the next phase of LIGO upgrades after O4 (`A+'), the goal is to achieve 6~dB of frequency-dependent squeezing improvement to the detector sensitivity \cite{Miller2015, Barsotti2018, McCuller2021}.
The piezo-deformable mirrors developed here have critical applications in active wavefront control for future gravitational-wave detectors, like Cosmic Explorer~\cite{CEHS}, as frequency-dependent squeezing is integral in achieving their design sensitivity.

A common architecture of piezo-deformable mirrors utilizes a thin mirror bonded with a piezoelectric substrate, referred to as the unimorph~\cite{roddier_1999, tokovinin2004using, rausch2016unimorph}.
The range of deformation is inversely proportional to the flexure rigidity, hence thinner unimorph mirrors yield a larger range of deformation.
Current unimorph technology does not simultaneously meet the reflectivity, surface quality, actuation range, and low defocus noise requirements for mode-matching optics for aLIGO, discussed in section~\ref{sec:des_req}.

Presented here is the design and implementation of a new deformable mirror for active mode matching in aLIGO.
The design uses a piezoelectric transducer (PZT) to apply a distributed bending moment on the mirror barrel via an axisymmetric flexure, thereby controlling the mirror radius of curvature.
The flexure-based piezo-deformable mirrors presented here has a large operating range, a high bandwidth, and is compatible with ultra-high vacuum operation.
Designed for two-inch diameter optics with 5~mm thickness, the scattering to higher-order modes is below 0.2$\%$ for beam radius less than 2~mm.
An alternative thermally actuated design is being implemented in aLIGO concurrently with the piezo-deformable mirrors \cite{Cao20}.
The thermal design has an increased operating range in vacuum compared to the piezo-deformable mirrors but offer a much lower bandwidth of $\mathcal{O}$(1~mHz) compared $\mathcal{O}$(1~Hz) that is achievable with the piezo-deformable mirrors.

The motivation and design requirements of the piezo-deformable mirrors for application in aLIGO and future upgrades are discussed in section~\ref{sec:des_req}.
In section~\ref{sec:design} the design of the flexure-based piezo-deformable mirror is presented.
The performance results of the piezo-deformable mirror are summarized in section~\ref{sec:res}.
%Section~\ref{sec:assm} describes the assembly procedure for the piezo-deformable mirrors.

\begin{figure}
    \begin{center}
    \includegraphics[width=\captionwidth]{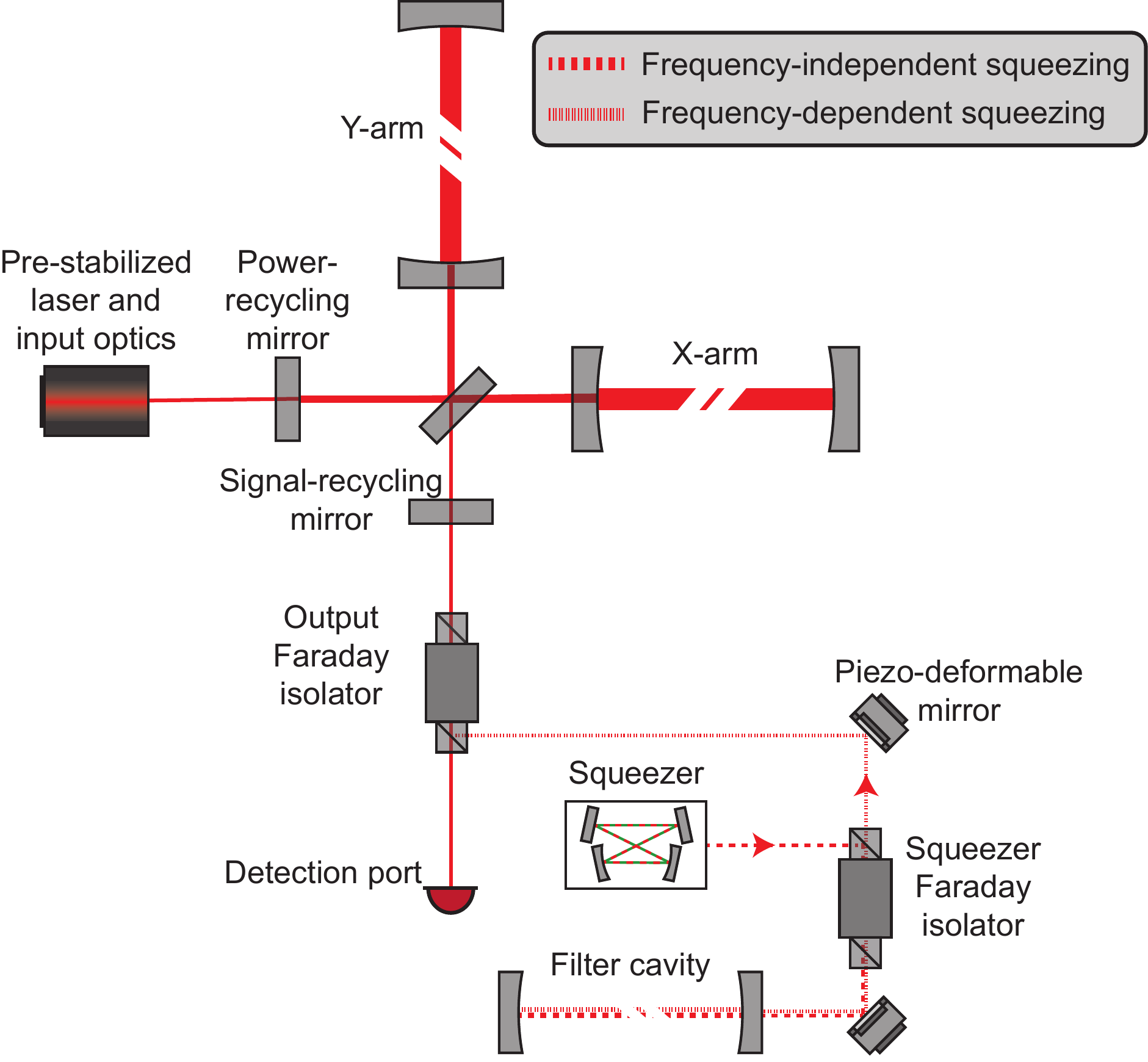}
    \caption{Simplified layout of the aLIGO detector with piezo-deformable mirrors in the squeezer path.
    In the aLIGO layout for O4, there is one piezo-deformable mirror between the squeezer and the filter cavity, and two piezo-deformable mirrors between the squeezer and output Faraday isolator. Only one is shown above for simplicity.}
    \label{fig:layout}
    \end{center}
    \end{figure}

%%%%%%%%%%%%%%%%%%%%%%%%%%%%%%%%%%%%%%%%%%%%%%%%%%%%%%%%%%%%%%%%%%%%%%%%%%%%%%%%%%%%%%%%%%%%%%%%%%%%%%
%%%% Section II
%%%%%%%%%%%%%%%%%%%%%%%%%%%%%%%%%%%%%%%%%%%%%%%%%%%%%%%%%%%%%%%%%%%%%%%%%%%%%%%%%%%%%%%%%%%%%%%%%%%%%%
\section{Requirements for aLIGO}\label{sec:des_req}
The piezo-deformable mirrors will be used for mode-matching at two locations in the aLIGO beam path, shown in Fig.~\ref{fig:layout}.
First, between the squeezed light source and the filter cavity. 
Second, between the output of the filter cavity and the main interferometer.
The required operating range of the piezo-deformable mirror is greater on the filter cavity to interferometer path, as the mode shape of the beam from the interferometer is not as well known.
However, for both of these ports, the mode-matching needs to be better than 96$\%$ to achieve the broadband improvements in sensitivity from 6~dB of frequency-dependent squeezing.
Based on our simulations of the current and expected mode mismatch, a range of $\pm160$~mD will be sufficient to correct for mode mismatch between the filter cavity and interferometer, and $\pm30$~mD is required on the path between the squeezer and the filter cavity. %T2000049
During future upgrades, to assist in characterizing the mode mismatch between the cavities in the aLIGO detector, the radius of curvature of the piezo-deformable mirrors can be dithered at a frequency in the aLIGO detection band~\cite{O3comm}. 
To achieve this, a high bandwidth $\mathcal{O}$(1-10~Hz) is needed on the piezo-deformable mirror actuators.

The aLIGO detectors have stringent requirements for any new optics added to the system.
Any technical noise added to the gravitational-wave readout must be at least a factor of 10 below the design sensitivity.
The piezo-deformable mirrors could inject technical noise through spurious changes in the defocus of the beam, higher-order mode content generated by the mirror surface, or displacement of the mirror surface. 
Any modes other than the fundamental Gaussian mode are considered to be higher-order modes.
We require the higher-order modes induced by the piezo-deformable mirrors to be a factor of 10 below the mode-mismatch ($L_{\text{o}}$) requirement of 4$\%$.
This demands that the higher-order mode content (or scattering) from the piezo-deformable mirrors be less than 0.4$\%$.

The defocus noise $S(f)$ arises from the fluctuations in the radius of curvature of the piezo-deformable mirrors.
This induces a fluctuation in the mode-matching losses, which produces a proportional change in relative intensity noise transmitted to the detection port, resulting in an apparent displacement noise ($z(f)$).
This apparent displacement noise for a given optic should be at least 10 times less than the displacement noise requirement in the aLIGO detectors so that it is not limiting the gravitational-wave detector sensitivity.
The defocus noise is dependent on the beam size ($w$) at the piezo-deformable mirror, and is coupled to the displacement noise by
\begin{equation}
    \frac{z(f)}{\mathcal{C}_{\text{tf}}} \approx \frac{\pi w^2}{\lambda}\sqrt{L_{\text{o}}} S(f),
\end{equation}
where $\mathcal{C}_{\text{tf}}$ is the transfer function from relative intensity noise to interferometer displacement noise, $\lambda$ is the wavelength of the laser, and $L_{\text{o}}$ is the dc mode matching requirement described above.
An analogous coupling due to the relative intensity noise requirement of the coherent locking field to the detection port sets the on the defocus noise requirement in the filter cavity path~\cite{McCuller_FC2018}, shown in Fig~\ref{fig:layout}.
Using this estimate $S(f)$ must be below $10^{-5}~\text{D}/\sqrt{\text{Hz}}$ above 100~Hz to meet the displacement noise requirement of the squeezer path.
This is a conservative estimate because at the optimally tuned setting of the piezo deformable mirror the linear coupling from defocus noise to displacement noise goes to zero.
Lastly, to damp displacement noise, and to soften the scattered light requirements, the piezo-deformable mirrors will be hung from double pendular suspensions, in a similar configuration to other aLIGO auxiliary optics~\cite{strain2012damping}.

%%%%%%%%%%%%%%%%%%%%%%%%%%%%%%%%%%%%%%%%%%%%%%%%%%%%%%%%%%%%%%%%%%%%%%%%%%%%%%%%%%%%%%%%%%%%%%%%%%%%%%
%%%% Section III
%%%%%%%%%%%%%%%%%%%%%%%%%%%%%%%%%%%%%%%%%%%%%%%%%%%%%%%%%%%%%%%%%%%%%%%%%%%%%%%%%%%%%%%%%%%%%%%%%%%%%%
\section{Design}\label{sec:design}

\begin{figure}
\begin{center}
\includegraphics[width=0.38\columnwidth]{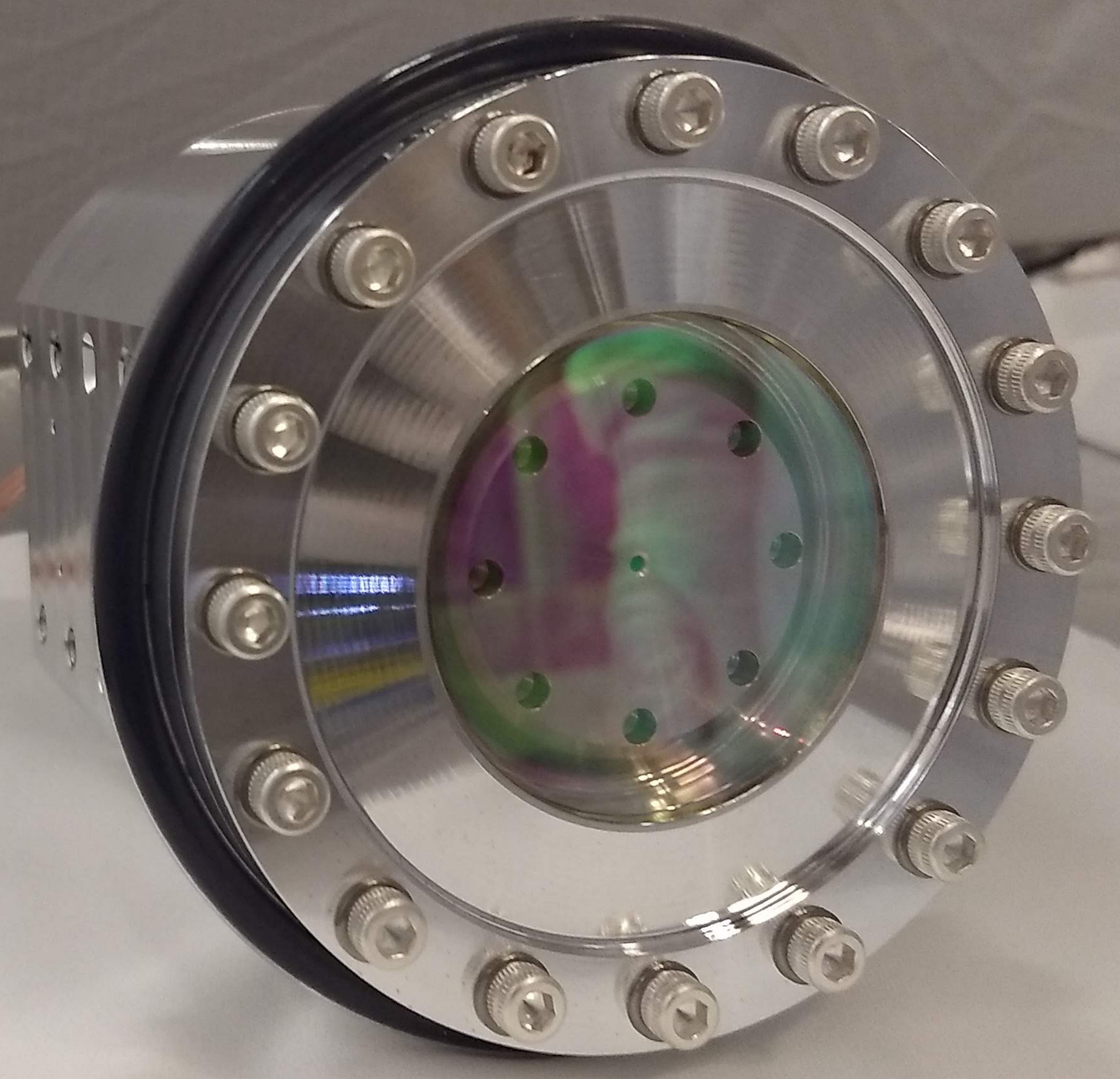}
\includegraphics[width=0.6\columnwidth]{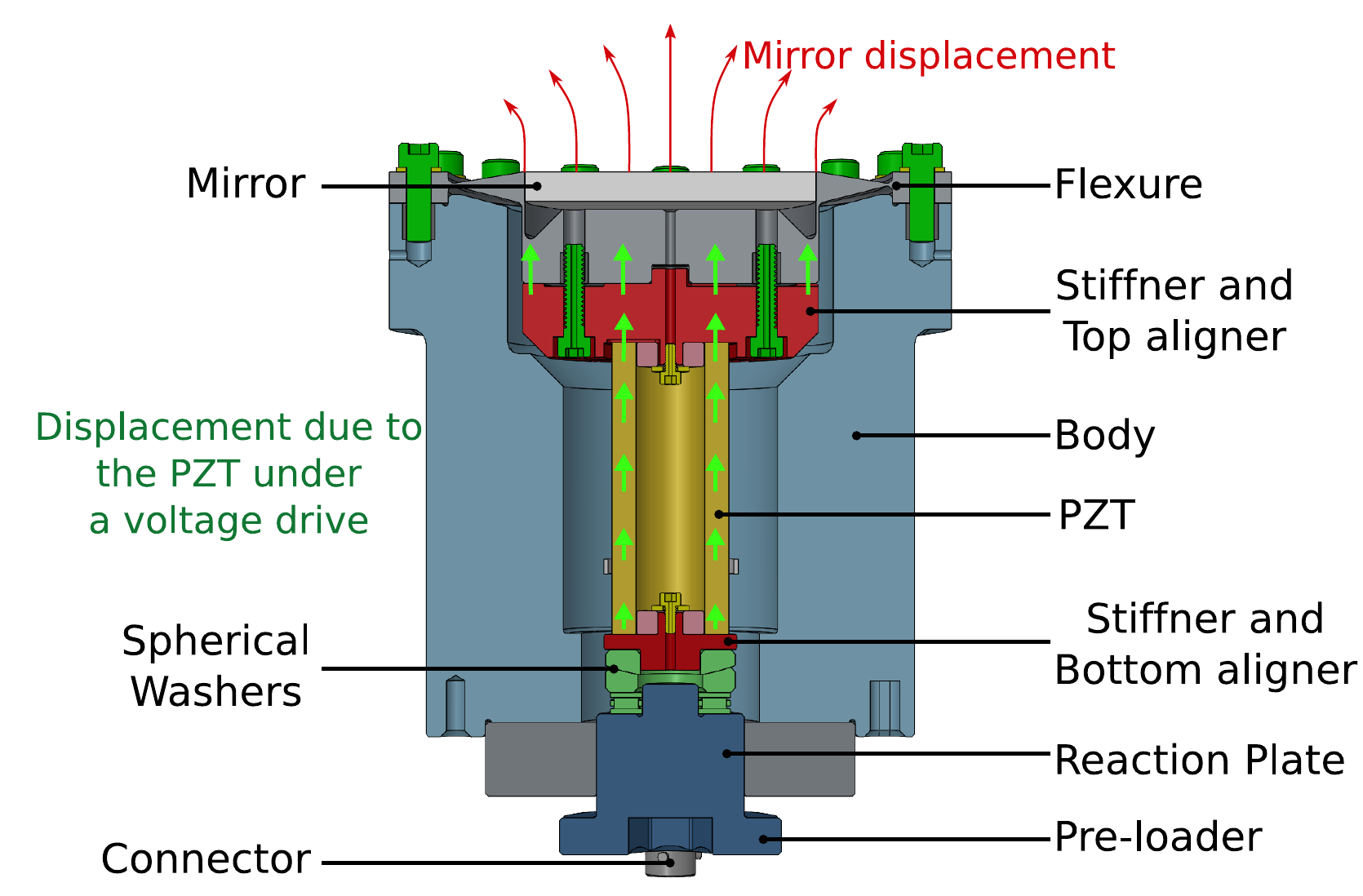}
\caption{\textit{Left:} An assembled flexure-based piezo-deformable mirror actuator for active mode matching in aLIGO.
\textit{Right:} The main components of an assembled piezo-deformable mirror. 
The mirror is compression fitted inside the flexure~\S\ref{subsec:comp_fit} and attached to the top aligner.
This assembly is then attached to the body using 16 screws each torqued to 30~in-lbs. 
The piezoelectric transducer (PZT) stack, and the bottom aligner are inserted from the back, followed with the attachment of the reaction plate.
The spherical washers and the thrust bearing are gently placed on the bottom aligner, and the pre-loader is torqued to at least 25~in-lbs to secure the assembly. 
The displacement of the PZT under a voltage bias is shown with green arrows. 
The flexure deforms the mirror radius of curvature due to the corresponding axial force generated by the PZT thrust. The mirror displacement is represented by red arrows.
}
\label{fig:PDM-Design}
\end{center}
\end{figure}

The piezo-deformable mirror is designed to meet the noise requirements described in section \ref{sec:des_req}, while maintaining a fast response time for convenient commissioning.
Fig.~\ref{fig:PDM-Design} shows the schematic of the piezo-deformable mirror. 
The key component is an inverted hat-shape axisymmetric flexure that converts a pushing force from the PZT into a distributed bending moment around the mirror circumference, thereby deforming the mirror for a radius of curvature.
When a voltage is applied to the PZT, it elongates along the axial direction.
The force associated with this elongation is distributed on the back of the flexure via the top aligner.
The flexure is bolted to the body to constrain any motion along the circumference, and the axial force applied to the flexure produces a moment that causes spherical deformation on the mirror.
Thus, by driving a voltage to the PZT one can actively change the radius of curvature of the mirror for mode-matching applications in aLIGO. 
The different components of the piezo-deformable mirror, see Fig.~\ref{fig:PDM-Design}, along with their functions are described below.
\begin{itemize}
  \item The {\it flexure} converts the axial force applied on the back to spherical deformation of the mirror surface. The mirror is held in the flexure due to the compression bias introduced after compression fitting.
  \item The radius of curvature of a 5~mm thick highly-reflective mirror is deformed for active wavefront control. These mirrors require a good barrel finish to minimize astigmatism, spherical aberrations, and coma which may arise from poor compression fitting.
  The process for compression fitting the mirror inside the flexure is discussed in section~\ref{subsec:comp_fit}.
  \item The flexure with the compression fitted mirror is attached to the {\it body} with screws. To ensure radial symmetry each screw is torqued to 30~in-lbs.
  \item The PZT (Noliac NAC2125-H50-A02) is held inside the {\it body} with {\it top and bottom aligners}. These aligners ensure that the piezo is flush against the back surface of the flexure. It is important to ensure that there is no angular misalignment of the PZT upon assembly. The PZT has a half-bridge strain gauge bonded to it, which allows the read out of the strain on the piezo, discussed in section~\ref{subsec:strain_gauge}.
  \item The {\it reaction plate} serves the role of a hard boundary wall. The applied force at the back of the flexure depends linearly on the longitudinal displacement. The reaction plate ensures that the stroke from the PZT preferentially displaces the back surface of the flexure.
  \item The {\it pre-loader} is a fine threaded screw that goes through the reaction plate. The pre-loader ensures that the piezo is secured stiffly inside the body. We recommend that the pre-loader is torqued to at least 25~in-lbs to ensure the PZT stack is well constrained before the application of any voltage drive. The torque applied to the pre-loader provides static deformation to the radius of curvature of the mirror without any voltage drive to the PZT. The operating point or the optimal radius of curvature can be changed by further torquing the pre-loader. One can torque the pre-loader up to 100~in-lbs without any damage to the piezo or the mirror. In this design, the pre-loader torque cannot be adjusted under vacuum.
\end{itemize}

The piezo-deformable mirror offers two design variations for the assembly of the mirror and the flexure.
First, a stainless steel (440C) flexure with an intermediate aluminum ring (6061) between the flexure and the mirror.
The design concept is presented in~\cite{ASPE2020}.
This allows greater flexibility in design because the level of mirror compression fit can be adjusted by compensating accordingly the intermediate ring.
However, the Young's modulus of stainless steel is higher, which limits the actuation range of the piezo-deformable mirror.
Also, the two-stage compression fitting -- flexure and ring, followed by mirror in the flexure -- makes compression fitting more challenging.
The second variation of the design uses an aluminum (7075) flexure which is custom made for each mirror.
This design, although tuned for each flexure-mirror pair, is easier to assemble.
Moreover, the lower Young's modulus of aluminum provides a greater actuation range as it is easier to deform the flexure.
We will reserve all discussions to the aluminum flexure-based piezo-deformable mirror in this paper.

Compression fitting the mirror into the flexure and installing the flexure assembly on the body induce minimal change to the defocus of the mirror.
The defocus is defined as the inverse of the focal length.
The net defocus of the piezo-deformable mirror ($D_{\text{PDM}}$) is given by 
\begin{equation}
D_{\text{PDM}} \approx D_{\text{mirror}} + D_{\text{preload}} + D_{\text{actuation}}
\end{equation}
where $D_{\text{mirror}}$ is the defocus of the mirror, $D_{\text{preload}}$ is the defocus due to the preload, and $D_{\text{actuation}}$ is the defocus due to the PZT actuation. The operating defocus $D_{\text{op}}$ is set such that $D_{\text{mirror}}$ and $D_{\text{preload}}$ cancel out at half the maximum actuation voltage.
\begin{equation}
\begin{split}
    D_{\text{op}} 
    & = D_{\text{mirror}} + D_{\text{preload}} + D_{\text{actuation}} (V_{\text{max}}/2) \\
    %& \approx D_{\text{actuation}} (V_{\text{max}}/2) \\
\end{split}
\end{equation}
The operating defocus for production units to be installed in aLIGO varies depending on the optic placement along the beam.

%%%%%%%%%%%%%%%%%%%%%%%%%%%%%%%%%%%%%%%%%%%%%%%%%%%%%%%%%%%%%%%%%%%%%%%%%%%%%%%%%%%%%%%%%%%%%%%%%%%%%%
%%%% Section V
%%%%%%%%%%%%%%%%%%%%%%%%%%%%%%%%%%%%%%%%%%%%%%%%%%%%%%%%%%%%%%%%%%%%%%%%%%%%%%%%%%%%%%%%%%%%%%%%%%%%%%
\section{Results}\label{sec:res}
Multiple assemblies of the piezo-deformable mirrors were tested using the Zygo interferometer, which is used to characterize the aLIGO core optics~\cite{zygo}.
The Zygo interferometer is a Fizeau topology, which uses a reference optic to measure the surface profile of a mirror under test.
We use the Zygo interferometer to measure the deformed mirror surface of the piezo-deformable mirror with varying amounts of torque applied to the pre-loader, and when the piezo is driven with an external voltage.
The measured deformation of the mirror was used to estimate the range of defocus and the higher-order mode content, which is discussed in sections~\S\ref{subsec:range} and~\S\ref{subsec:HOMS}.
The defocus noise of the piezo-deformable mirror has also been measured in a separate Michelson interferometer and the preliminary results suggest that aLIGO requirements along the squeezer path are met by almost four orders of magnitude~\cite{Jia2021}.
As part of the Michelson interferometer testing, the resonances of the piezo-deformable mirror assembly have been measured, with the lowest resonance at 386 Hz.
With a 100~V bias on the PZT, we measure a bandwidth of 6.8~Hz for the piezo-deformable mirror.
This bandwidth is limited by the driver electronics.
During O4, the piezo-deformable mirror will be tuned occasionally to reduce mode-matching losses.
In future upgrades, the fast response of the piezo-deformable mirrors allows the possibility to design feedback control loops to reduce mode-mismatch losses.
 
\subsection{Defocus range of the piezo-deformable mirror}\label{subsec:range}
\begin{figure}
\begin{center}
\includegraphics[width=\captionwidth]{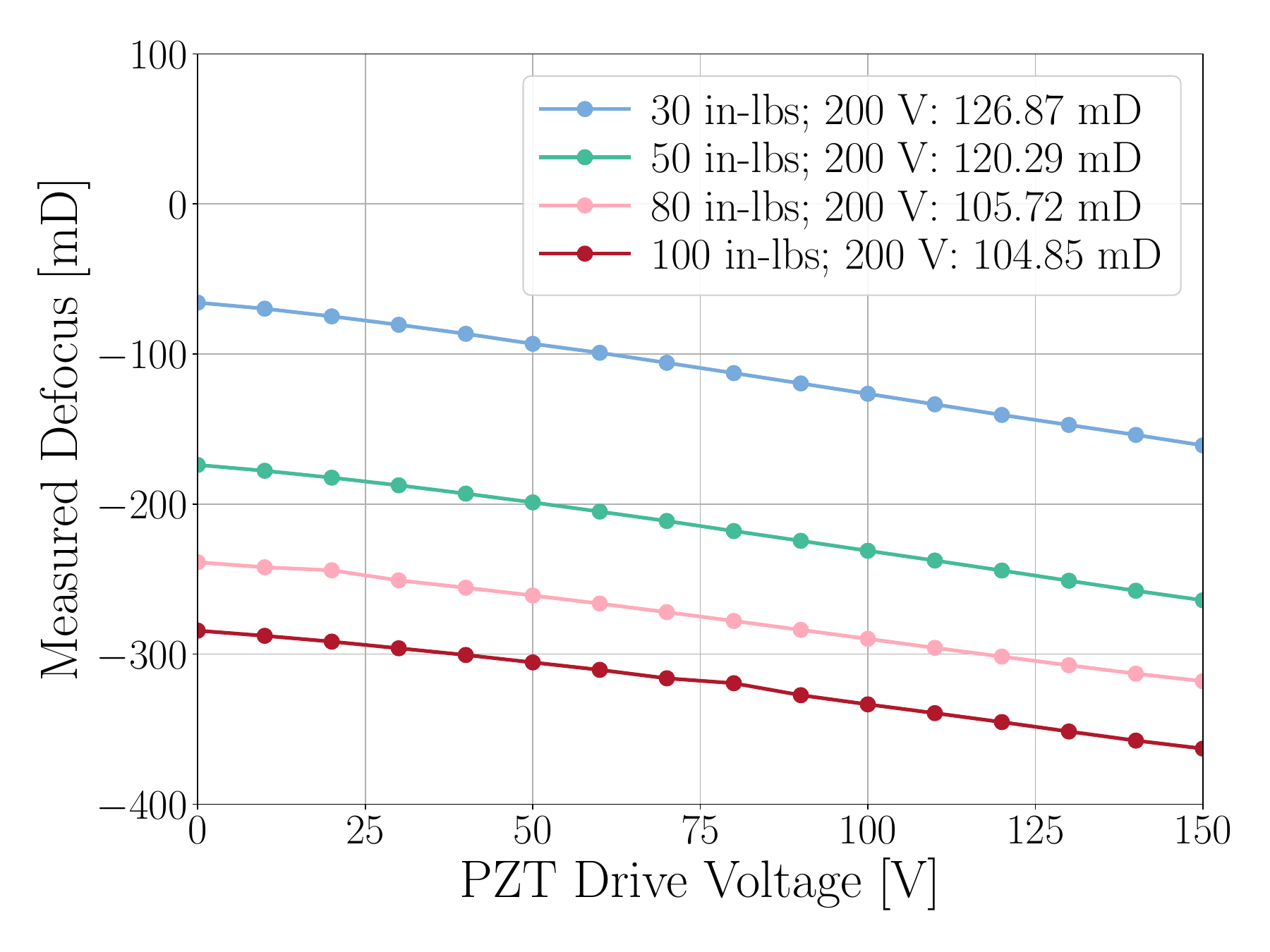}
\caption{The points represent the measured defocus range (mD) as a function of drive voltages through the PZT,
with a static pre-loads of 30, 50, 80 and 100 in-lbs (blue, green, pink and red). 
The legend summarizes the total defocus range extrapolated over a 200~V drive at the corresponding static pre-loads.
We note that the minimal 25~in-lbs preload induces an approximately 60~mD of defocus on the mirror. 
The static preload by adjusting the torque applied to the pre-loader (no voltage drive) offers a defocus range of over 200~mD 
(from -65~mD to -285~mD for this prototype).
The PZT drive over 200~V offers an additional \fullrange~mD of operating range.}
\label{fig:5mm-Al-Range}
\end{center}
\end{figure}

The piezo-deformable mirror actively changes the radius of curvature of the mirror when an external voltage is applied.
We use the Zygo interferometer to measure the surface profile of the mirror in the assembled piezo-deformable mirror~\cite{zygo}.
At different values of the piezo drive (from 0~V to 200~V), we measure the surface profile with a pixel resolution of 192~$\mu$m and with 100 averages.
The surface profile was fit to linear and quadratic order over a circular region with a diameter of 25~mm. 
The quadratic term measures the defocus and the linear term measures the tilt of the mirror surface.
Averaging the measured operating range from three different piezo-deformable mirror assemblies, each tested at different values of pre-loader torques, the piezo-deformable mirror offer an active operating range of \quarterrange~mD over 150~V of an external voltage.
The PZT in the design can be driven up to 200~V offering an operating range of \fullrange~mD.
The pre-loader was torqued to different values to change the static defocus of the mirror.
We find the change in preload from the minimum of 25~in-lbs to a maximum of 100~in-lbs offers approximately \staticrange~mD of static adjustment range, see Fig.~\ref{fig:5mm-Al-Range}.

\subsection{Higher-order mode scattering}\label{subsec:HOMS}
\begin{figure}
\begin{center}
\includegraphics[width=0.49\textwidth]{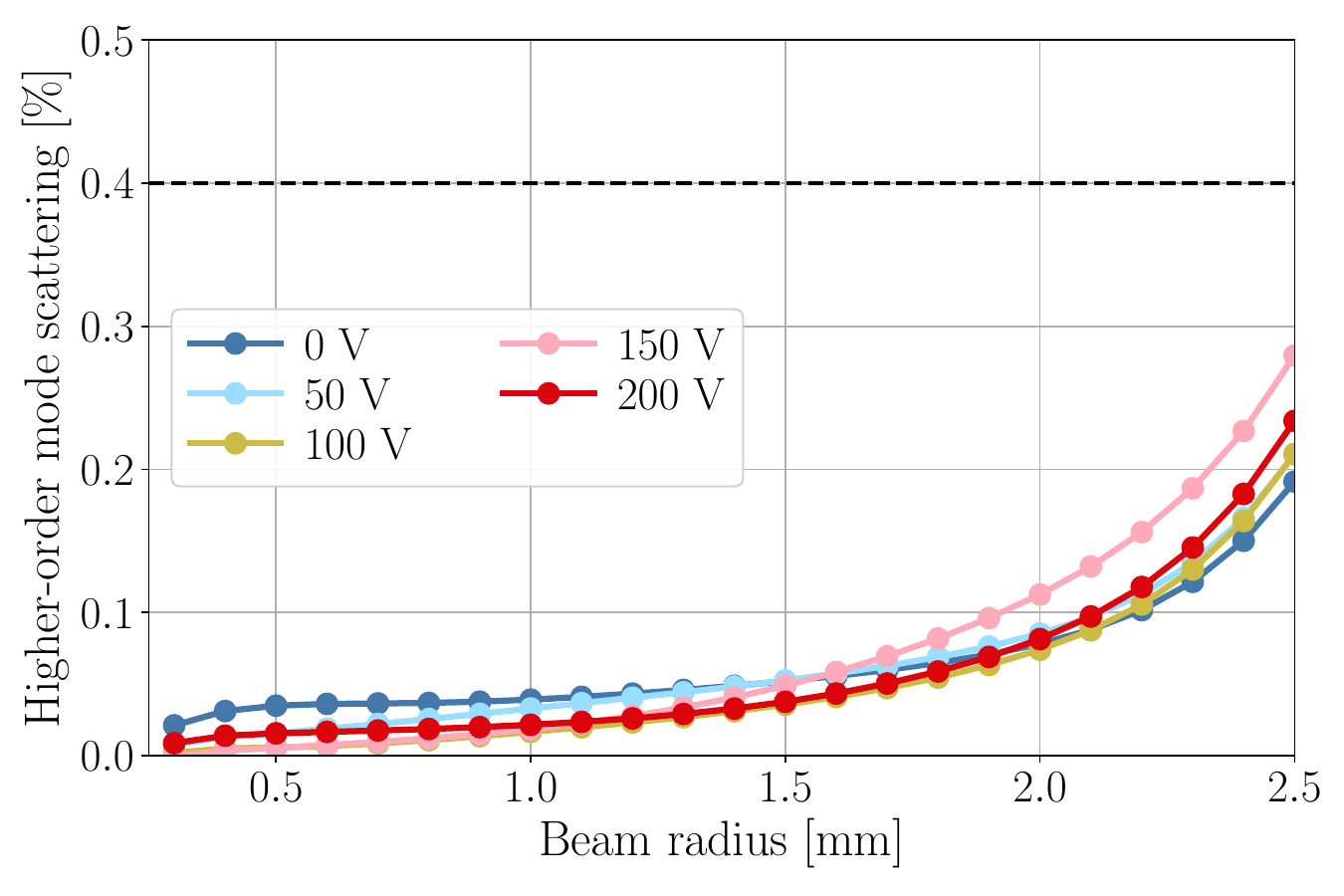} 
\includegraphics[width=0.49\textwidth]{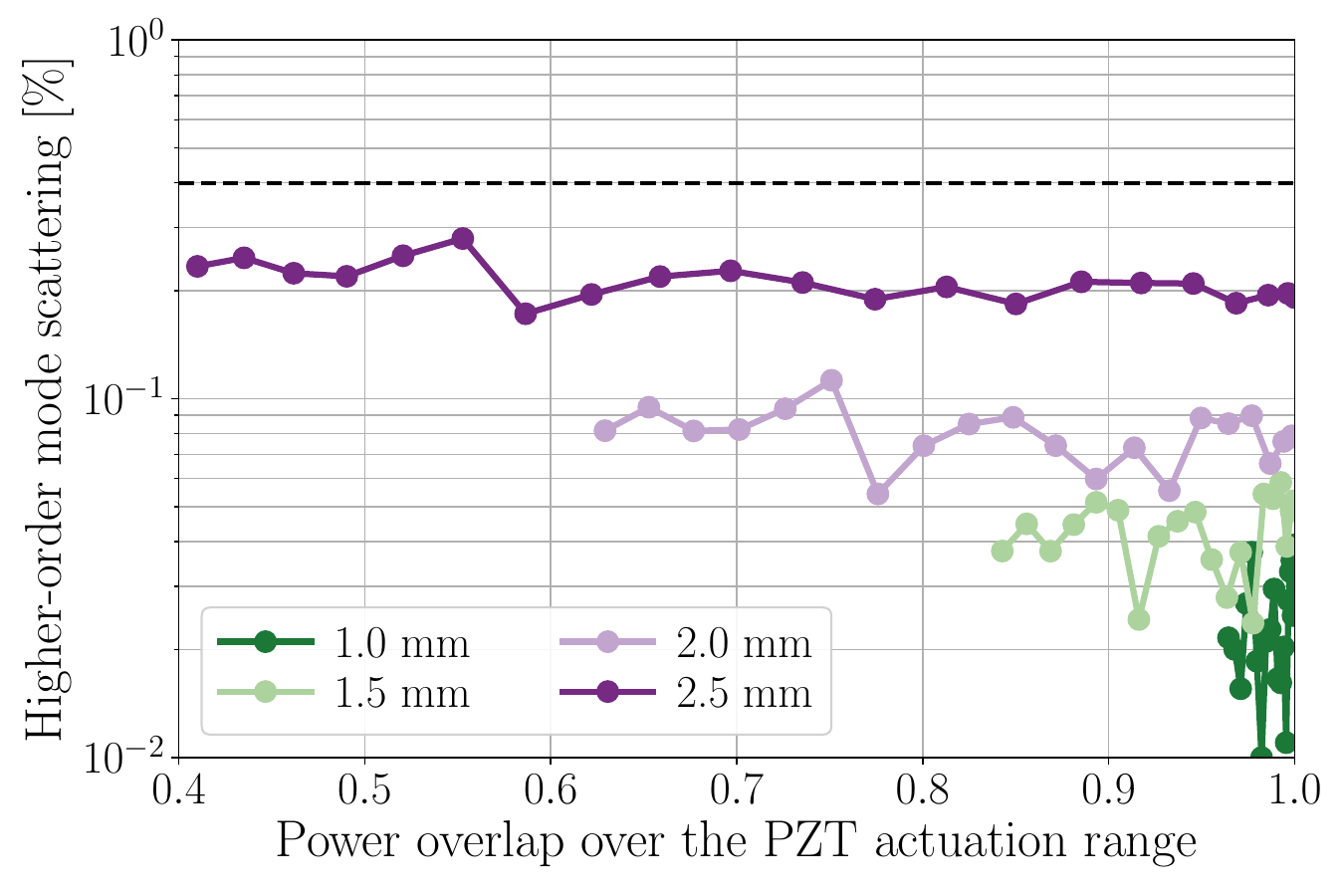}
\caption{\textit{Left:} The higher-order mode content of piezo-deformable mirrors with 50 in-lbs of preload at 0, 50, 100, 150, and 200~V of drive to the PZT as a function of beam radius. 
The Zygo interferometer has a pixel resolution of 0.192~mm, which limits the higher-order mode content projection for smaller beam sizes. 
However, it is expected to be smaller than the higher-order mode content for 0.25~mm beam radius.
We find Gaussian beams up to a radius of 2.5~mm experience less than 0.4$\%$ of higher-order mode content over the entire range of piezo-deformable mirrors operation.
\textit{Right:} The higher-order mode content as a function of power overlap between the incident and the reflected field.
As the defocus of the mirror increases with the PZT drive, the power overlap decreases.
However, we find the piezo-deformable mirrors ensures higher-order mode content less than 0.4$\%$ for Gaussian beams with beam radius up to 2.5~mm over the full actuation range of the PZT.
}
\label{fig:HOMS}
\end{center}
\end{figure}

The design specification requires the piezo-deformable mirror must induce less than 0.4$\%$ of higher-order mode power for Gaussian beams with a beam radius of less than 2~mm, see section~\ref{sec:des_req}.
To estimate the higher-order mode content we define a nominal reflected beam using an ideal Gaussian input beam with a given beam size. The input beam is reflected off a simulated mirror with defocus and tilt as measured by the Zygo (see Fig~\ref{fig:5mm-Al-Range}) but with no additional higher order mode content.
We calculate the full reflected beam using the measured surface profile and the same ideal Gaussian input beam.
The mode-overlap between the nominal and the full reflected beams provides an estimate of the higher-order mode content of the full reflected beam.
The left plot of Fig.~\ref{fig:HOMS} shows the higher-order mode content as a function of input beam radius over the range of the piezo-deformable mirrors. 
Alternatively, we can quantify the power-overlap between the input beam and the nominal reflected beam, correcting for pointing errors over the operating range.
The right plot of Fig.~\ref{fig:HOMS} shows that the piezo-deformable mirror provides low higher-order mode content 
even when correcting for an overlap mismatch as low as 0.4, demonstrating that the design constraints are met.

The compression fitting scheme discussed in section~\S\ref{subsec:comp_fit} is critical to achieving low higher-order mode content.
It is crucial to ensure during the process of compression fitting that the mirror has no tilt with respect to the flexure.
Any tilt between the mirror and the flexure causes a non-axisymmetric deformation of the mirror after compression fitting.
When the mirror is actuated in this configuration, the moment distribution on the mirror circumference generated by the flexure is not axially symmetric, which causes the higher-order mode content to be much higher, typically up to 2-5$\%$.

%%%%%%%%%%%%%%%%%%%%%%%%%%%%%%%%%%%%%%%%%%%%%%%%%%%%%%%%%%%%%%%%%%%%%%%%%%%%%%%%%%%%%%%%%%%%%%%%%%%%%%
%%%% Section VI
%%%%%%%%%%%%%%%%%%%%%%%%%%%%%%%%%%%%%%%%%%%%%%%%%%%%%%%%%%%%%%%%%%%%%%%%%%%%%%%%%%%%%%%%%%%%%%%%%%%%%%
\section{Conclusion}\label{sec:concl}
We present a novel ultra-high vacuum compatible, flexure-based active mode-matching deformable mirror.
The piezo-deformable mirrors presented here have direct implications for improving the sensitivity of aLIGO detectors by reducing the optical losses from mode-mismatch, and by improving the levels (dB) of squeezing.
They offer high bandwidth and a large operating range. While under vacuum, the PZT actuator has an operating range of \fullrange~mD. 
In air, adjusting the static preload by using the pre-loader provides an additional \staticrange~mD of adjustment range. 
The quality of the beam is not degraded and the higher-order mode content is below 0.2$\%$ over the full range of actuation.
The technology developed here has applications in any optical experiments where mode-matching is critical or active wavefront control is necessary.

\section{Acknowledgment}\label{sec:ack}
LIGO was constructed by the California Institute of Technology and Massachusetts Institute of Technology with funding from the National Science Foundation, and operates under cooperative agreement PHY-1764464. A+ was built under award PHY-1234382. This paper carries LIGO Document Number LIGO P2100315. Parts of this research were conducted by the Australian Research Council Centre of Excellence for Gravitational Wave Discovery (OzGrav), through project number CE170100004.

\section{Disclosures}
The authors declare no conflict of interest. 

\section{Appendix}

\subsection{Compression fitting and assembling the piezo-deformable mirror}\label{subsec:comp_fit}
In this section, we discuss the procedure to compress fit the mirror inside the aluminum flexure.
To induce compression bias at room temperature, the diameter of the mirror is larger than the inner diameter of the flexure.
Compression fitting a mirror for adaptive optics was also demonstrated in~\cite{Cao20}.
The procedure to compression fit the mirror is as follows:

\begin{itemize}
  \item As the coefficient of thermal expansion of aluminum are higher than fused silica, we can heat the flexure to create clearance to insert the mirror inside. The temperature to perform the compression fit is determined by the following equation
    \begin{equation}\label{eq:temp}
        \Delta T = T_{\text{fit}} - 300~K \approx \frac{\delta \phi}{\phi} \frac{1}{\alpha_{\text{flex}}}
    \end{equation}
  where $\alpha_{\text{flex}}$ is the coefficient of thermal expansion of the flexure, $\phi$ is the diameter of the mirror at 300~K and $\delta \phi$ is the interference between the mirror and the flexure at 300~K, typically 15~$\mu m$. The temperature $T_{\text{fit}}$ is the approximate temperature to perform the compression fit.
  \item We use the optical grade surface plate to ensure that the normal from the surface of the mirror is parallel to the normal axis of the flexure plane -- ensuring the mirror does not tilt inside the flexure. The mirror, the flexure, and the alignment tool are heated to $T_{\text{fit}}$.
  To minimize the non-radial deformation of the mirror after compression fit, the barrel of the mirror requires a good surface finish and cylindricity.
  The prototypes presented here used mirrors with cylindricity~$<$~5$\mu m$ (Thorlabs BB2-E03-5MMT-SP).
  The piezo-deformable mirrors employed at aLIGO use mirrors with specified cylindricity~$<$~1$\mu m$ (FiveNine Optics).
  \item At room temperature, place the mirror on the optical grade surface plate, align the flexure face-down such that it just makes contact with the back surface of the mirror along the circumference. Place a weight gently on top of the setup. Next, the temperature is increased in steps 15~K (to $T_{\text{fit}} \sim 550~K$) until the mirror is all the way inside the flexure. The assembly is then cooled to room temperature, ensuring no disturbance to the setup.
\end{itemize}
Next, the compression fitted flexure is first loosely attached to the body and the top aligner resting inside the body of the assembly.
With the body mirror-side-down, the PZT is engaged with the top aligner and bottom aligner via two nylon rings inserted inside the PZT.
A pair of spherical washers and thrust bearing are then placed over the bottom aligner, and the reaction plate is screwed down at the back of the body.
%The reaction plate has an internal thread so that the pre-loader is screwed from the back.
The pre-loader is then engaged with the bottom aligner via the spherical washers and thrust bearing.
Finally, all the screws in the flexure assembly need to be tightened.
All the screws to attach the compression fitted mirror-flexure assembly to the body are torqued to 30~in-lbs. 
The pre-loader is torqued to at least 25~in-lbs; the maximum allowable torque to the pre-loader, ensuring no damage to the mirror, is 100~in-lbs.

\subsection{Strain Gauge Readout}\label{subsec:strain_gauge}
\begin{figure}
\begin{center}
\includegraphics[width=\captionwidth]{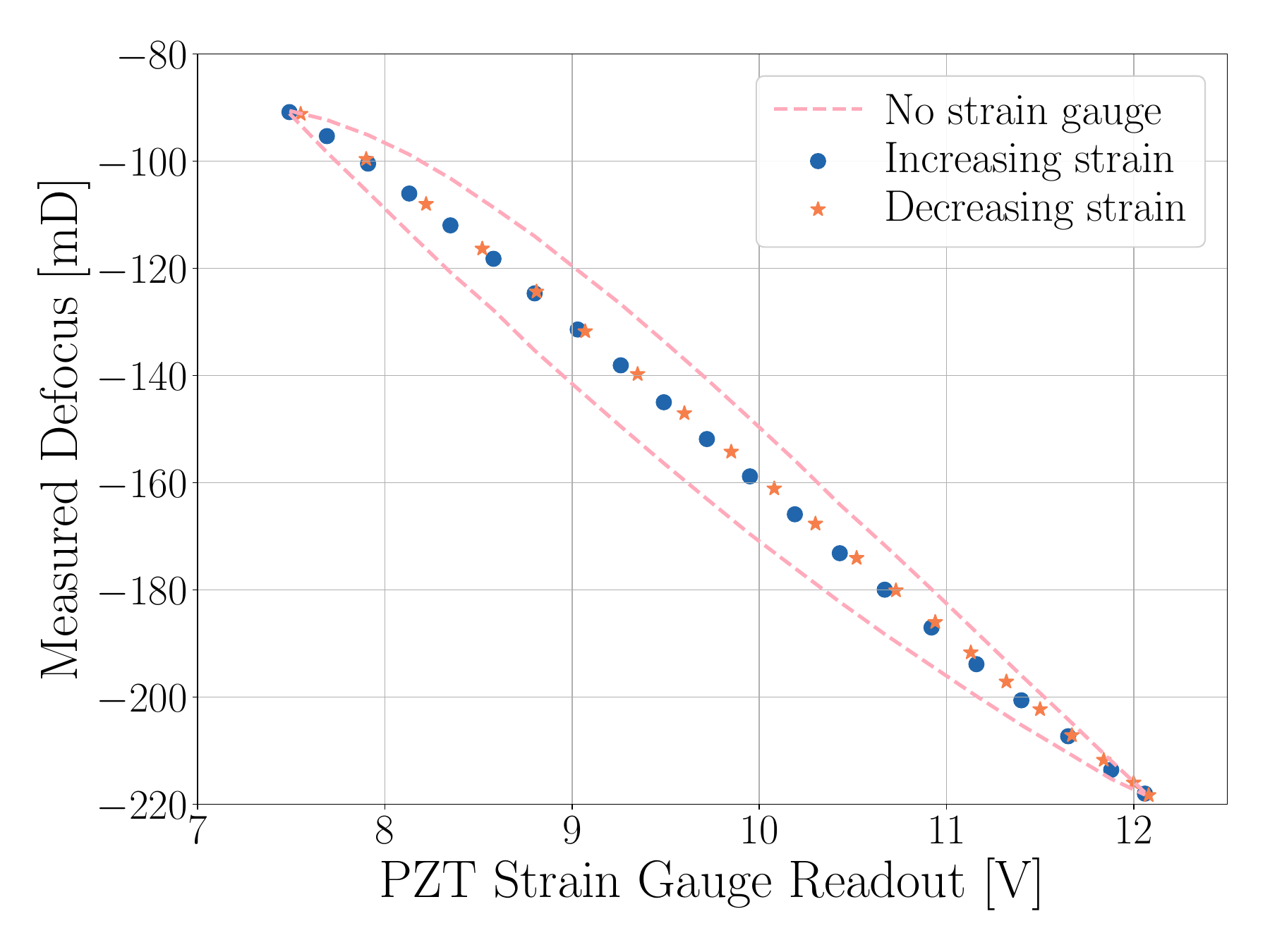}
\caption{The dashed pink outer boundary to the curve shows the measured defocus resulting from a 200~V peak-to-peak sinusoidal voltage applied to the piezo actuator.  The visible hysteretic response creates uncertainty if the drive voltage were used to define the resulting defocus. The blue circles (orange stars) represent the measured defocus at 10V steps of the sinusoidal drive voltage as a function of the strain-gauge readout bonded to the PZT. This illustrates how the strain gauge readout provides a linear and repeatable measurement of the defocus.  The linear readout feature in conjunction with the high-bandwidth of the piezo-deformable mirrors enables closed-loop mode-matching and active wavefront applications.
}
\label{fig:PZT_hys}
\end{center}
\end{figure}

The crystalline structure of the piezoelectric material gives rise to hysteresis in the actuation curve as seen in Fig~\ref{fig:PZT_hys}.
This hysteresis is undesirable for mode-matching applications requiring the ability to revert to a known radius of curvature while the piezo-deformable mirror is otherwise inaccessible inside an ultra-high vacuum environment. 
%The hysteresis is produced by remnant polarization within the piezoelectric crystal and leads to a residual longitudinal displacement when the applied PZT electric field is reversed.
The PZT in the piezo-deformable mirror is manufactured with an integrally bonded half-bridge strain gauge readout. 
The half-bridge is comprised of separate collocated transverse and axial strain gauges, orthogonally mounted with respect to each other on the barrel of the cylindrical PZT. 
The measurement of strain using this type of gauge relies on the measurement of small changes in resistance (1 ohm full-scale) associated with the elongation of the conductive elements in the sensor. 
An axially mounted strain gauge measures the elongation of the PZT as it is driven by an external voltage.
As the axial sensor is stretched and compressed, an undesired change in area results from the Poisson-ratio of the piezo-ceramic material much as would be seen by stretching a rubber band.
A second transversely mounted strain gauge provides compensation for the unwanted area change. 
A side benefit to the use of two collocated sensors is inherent temperature compensation to the resulting half bridge circuit. 
The combined results of these techniques in conjunction with a balanced Wheatstone bridge readout are linear and repeatable defocus measurements as shown in Fig~\ref{fig:PZT_hys}.

\bibliography{SAMS}

\end{document}